# Identification of two-dimensional FeO$_2$ termination of hematite α-Fe$_2$O$_3$(0001) surface

*Jesús Redondo*[*,1], *Petr Lazar*[3], *Pavel Procházka*[4], *Stanislav Průša*[4], *Jan Lachnitt*[2], *Benjamín Mallada*[1], *Aleš Cahlík*[1], *Jan Berger*[1], *Břetislav Šmíd*[4], *Pavel Jelínek*[1], *Jan Čechal*[*,4], *Martin Švec*[*,1]*

1 - Institute of Physics, Czech Academy of Sciences, Prague, Czech Republic
2 - Faculty of Mathematics and Physics, Charles University, Prague, Czech Republic
3 - Regional Center for Advanced Materials and Technologies, Olomouc, Czech Republic
4 - Central European Institute of Technology, Brno, Czech Republic

## ABSTRACT

Iron oxides are among the most abundant compounds on Earth and have consequently been studied and used extensively in industrial processes. Despite these efforts, concrete understanding of some of their surface phase structures has remained elusive, in particular the oxidized α-Fe$_2$O$_3$(0001) hematite surface. We detail an optimized recipe to produce this phase over the entire hematite surface and study the geometrical parameters and composition of its complex structure by means of atomically resolved microscopy, electron diffraction and surface-sensitive spectroscopies. We conclude that the oxidized α-Fe$_2$O$_3$(0001) surface is terminated by a two-dimensional iron oxide with structure, lattice parameters, and orientation different from the bulk substrate. Using total-energy density functional theory for simulation of a large-scale atomic model, we identify the structure of the surface layer as antiferromagnetic, conductive 1T-FeO$_2$ attached on half-metal terminated bulk. The model succeeds in reproducing the characteristic modulations observed in the atomically resolved images and electron diffraction patterns.

Mineral iron oxide is known to be available in many stoichiometries, polymorphs, and even mixtures. Hematite (α-$Fe_2O_3$), maghemite (γ-$Fe_2O_3$), magnetite ($Fe_3O_4$), and wüstite ($Fe_{1-x}O$) are prominent representatives of this class, possessing a wide range of electronic, magnetic and catalytic properties, due to their different oxygen content and characteristic crystal structures [1]. Currently, a significant amount of research is focused on the catalytic processes that occur on the surfaces of iron oxides, such as CO oxidation, wastewater purification, liquid fuel synthesis via Fischer-Tropsch reactions, styrene production or water splitting [2].

The research on iron oxides has turned its focus to nanoparticles [3] and thin films, [4] which are economically advantageous and show versatility beyond that of bulk materials. The critical limit - 2D iron oxide films have been achieved in a form of a monolayer of FeO(111) on Pt(111) [5], Ag(111) [6], Ru(0001) [7] and Pd(111) [8]. Nevertheless, stability of this monolayer is inherently linked to its strong hybridization with the metal substrate; it remains an open question whether it can exist independently or as termination of an iron oxide crystal [9,10], in analogy with the recently revisited $V_2O_3$(0001) system [11,12]. Another recent survey on new possible candidates for 2D materials suggests that some trilayer structures of metal oxides (e.g. $MnO_2$, $CoO_2$, $GeO_2$) may be stable after exfoliation from a layered bulk [13].

The atomic structures of bulk $Fe_3O_4$ and α-$Fe_2O_3$ are well-known, but their surface terminations remain elusive [2,14], owing to their complexity and the fact that their surface stoichiometry can be varied depending on the preparation. Specifically, by removing oxygen from the surface of α-$Fe_2O_3$(0001) by selective sputtering, a stoichiometry and structure resembling $Fe_3O_4$(111) can be attained [15,16]. Conversely, the surfaces can be partially or fully reoxidized by increasing the oxygen chemical potential (pressure) during annealing in ultra-high vacuum (UHV) [2,17]. The resulting reconstructions display a characteristic nanoscale pattern manifested as a floreted low-energy electron diffraction (LEED). It has been initially interpreted as an FeO(111) overlayer on top of the bulk [18], or later explained by coexistence of alternating domains of FeO(111) and α-$Fe_2O_3$(0001), based on images taken by scanning tunneling microscopy (STM). Thereafter the name "biphase" has been coined for this particular surface reconstruction [19,20]. Whereas several works use the biphase concept to explain microscopy and spectroscopy data [15,21–26], other research somewhat counterintuitively suggests the existence of a thin $Fe_3O_4$(111) layer on the bulk,[27] or states that the surface is oxygen-terminated [28]. Moreover, despite a large number of density functional theory (DFT) calculations, currently there is no atomic model that comprehensively explains the various experimental observations [22,29–33].

In this work, we prepare a single-phase, fully oxidized surface of α-$Fe_2O_3$(0001), with the objective of proposing a model of its atomic structure using a multitude of microscopy and spectroscopy techniques, and rationalize it by total-energy DFT.

**Results and discussion**

In order to prepare the reduced phase of the surface (R-phase), the sample was sputtered with Ar[+] at 1 keV and subsequently annealed in UHV at 700 °C. Sputtering the surface selectively removes O atoms, which reduces the surface from $Fe_2O_3$ to stoichiometric $Fe_3O_4$ [15,34]. Long sputtering times are avoided as the reduction progresses into the bulk and its reoxidation poses a challenge [35]. A complete coverage of the substrate with the R-phase can be readily achieved with a few (3-5) 10 minute sputtering cycles. This surface phase is relatively flat, with irregular-shaped terraces typically over 100 nm wide - as shown by the STM in Fig. 1a, along with its characteristic 2×2 micro-LEED (µLEED) pattern consisting of two rotational domains mutually offset by 60°, distinguishable by taking diffraction patterns of separate domains. These domains can be also resolved in real-space by low-energy electron microscopy (LEEM), using a composition of dark-field images (orange and blue, also shown in Fig. 1a) obtained by imaging only electrons diffracted to a direction associated with a particular domain orientation. These domains are equally represented on the surface.

Upon annealing for 30 min in $1\times10^{-6}$ mbar of $O_2$ at 700 °C, the oxidized phase starts to grow (which we deliberately denote as *H-phase* for its characteristic honeycomb-like pattern, appearing in the STM shown in Fig. 2a). This level of re-oxidation is characterized by a heterogenous mixture of R- and H- phases, dominated by the R-phase; the STM reveals a morphologically complex landscape with a large number of small, atomically flat terraces with sizes typically around 50 nm (see Fig. 1b).

The final stage of the R- to H-phase transformation is accompanied by the growth of large single-domain terraces that often span over several hundreds of nm, typically divided by bunches of steps. It takes several hours of annealing in $O_2$ to fully transform the remaining R-phase into the H-phase, as indicated by complete disappearance of the 2×2 fractional spots in the diffraction pattern shown in Fig. 1c. The need for long annealing times well above 1h in $O_2$ to fully reoxidize the surface was already addressed by Saunders *et al.*, who observed that a single 30 min annealing at 700 °C led to no significant changes, on the other hand a significantly longer (90 min) annealing completely reoxidized the surface [34]. It has also been observed that the recovery of the $Fe_2O_3$ stoichiometry in the surface and subsurface region requires achieving the correct temperature [35].

In the course of the transition from the R- to the H-phase, the X-ray photoelectron spectroscopy (XPS) Fe 2*p* lineshape undergoes a characteristic evolution [36]. The appearance of satellite peaks at 732.8 and 719.0 eV indicates the development of the $Fe^{3+}$ oxidation state

(XPS spectra in Fig. 1b,c). However, using these spectra to quantify the progress of the transformation is problematic (or the Fe oxidation/spin state), mostly due to its notoriously complicated multiplet splitting [37].

Once the H-phase is formed, sample annealing in vacuum up to 900 ºC does not cause any reversal to the R-phase detectable by our methods. The surface covered by H-phase remains unaltered in UHV for several days; even after exposing it to air it is partially recoverable by in-vacuum annealing. In the latter case, we observed that a fraction of R-phase is formed upon re-insertion into UHV and annealing, which can be attributed to the effect of adventitious C being trapped from the atmosphere [38,39] (more details about the effect of air exposure and annealing in vacuum can be found in Fig. S1).

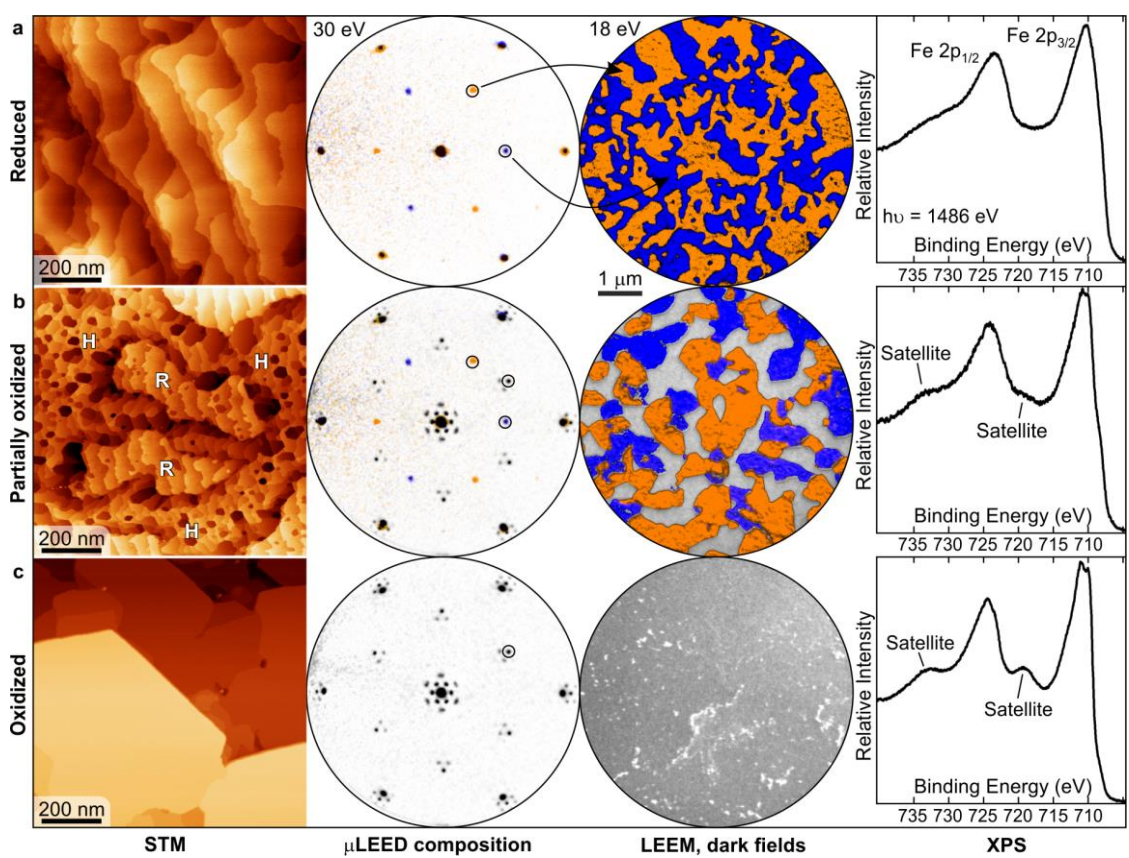

**Figure 1:** Oxidation process of the α-$Fe_2O_3$(0001) surface monitored by STM, μLEED, LEEM, and XPS (from left to right). a) Stoichiometric $Fe_3O_4$(111) surface obtained after sputtering and annealing in UHV. Colored μLEED and dark-field LEEM images reveal two rotational domains (shown in orange and blue). b) Partially oxidized surface after initial 30 min. annealing in $O_2$. Satellite peaks in the Fe 2*p* XPS corresponding to $Fe^{3+}$ begin to develop. c) Completely oxidized surface featuring large terraces.

Figure 2a shows a representative STM image of a smaller area of the H-phase surface displaying a superstructure on two terraces divided by a step edge (Δz = 0.48 ± 0.02nm). The superstructure visible on both terraces is hexagonal, resembling a honeycomb. The orientation of the higher and lower terrace honeycombs differs by an angle of 8 ± 1º. The average periodicity is 40 ± 1 Å. The spatial frequency and the angle of the mutual domain rotation can be corroborated by comparing the combined FFT power spectrum in Fig. 2b, derived from the STM image, with colors distinguishing the rotational domains, a µLEED pattern taken near the (00) spot (Fig. 2c), also combined from two single domain diffraction patterns taken on two rotational domains. These reciprocal space images are in good agreement, with the periodicity and the angle being reproduced.

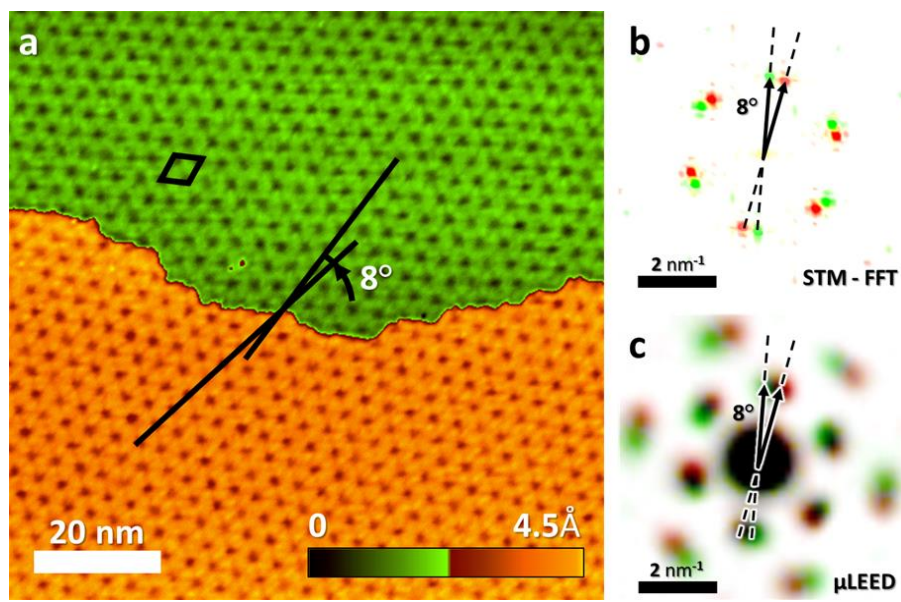

**Figure 2:** Characterization of the honeycomb superstructure. a) STM constant-current image, setpoint 0.3 nA, 1.4 V. The black lines mark the apparent domain orientations. b) FFT image combined from the upper terrace (green) and lower terrace (orange). c) µLEED of the (00) spot area, colored and composed from the patterns acquired from two adjacent terraces observed in LEEM, measured with 30 eV electron energy and 185 nm electron beam spot diameter. The dashed black lines denote the honeycomb domain orientations; the black arrows are their corresponding reciprocal supercell vectors.

The two atomically-resolved STM images in Figs. 3a,b show in detail a few supercells of the honeycomb superstructure. We would like to emphasize that the only difference between these two very distinct contrast types is the tip state; both types of contrast were regularly observed on the identical surface phase, and are independent of the bias voltage. Figure 3a shows a uniform layer with a periodicity of 3.08 ± 0.08 Å, intensity-modulated by the 40 Å honeycomb strikingly reminiscent of a moiré pattern. Figure 3b shows the same long-range modulation, but due to an emerging local √3×√3 periodicity. The latter type of contrast corresponds to previous works [19,20]; however our measurement provides higher resolution and proves that there are extra bright spots in the regions with the √3×√3 periodicity that are all within the same first 1×1 sublattice.

Figure 3c shows the LEED pattern associated with this surface, exhibiting the characteristic florets around the bulk diffraction spots [18,19]. The (01) and (11) spots originate from the bulk α-$Fe_2O_3$(0001) [40] unit cell and are taken as the reference ($a_{bulk}$ = 5.038 Å) [41] for the purpose of the superstructure analysis. The apparent angle of the superstructure orientation (denoted by a black vector on a dashed line in Fig. 3c with respect to the (10) bulk vector is $|\alpha|$ = 26° ± 0.5°. Notably, the most intense spots (marked by a cyan vector) are located within the florets around the (11) bulk spots. Their shape is tangentially elongated, since they comprise two individual spots. We determined the corresponding periodicity to be $a_{top}$ = 3.14 ± 0.01 Å, which reasonably matches the fine modulation observed via STM and its rotation with respect to the (11) vectors as $\delta$ = 0.31° ± 0.04°.

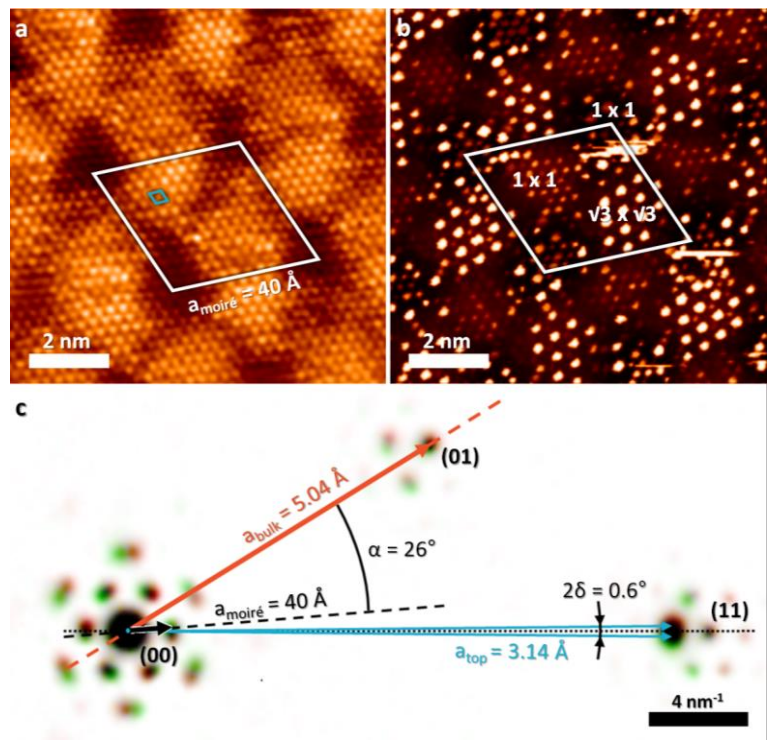

**Fig. 3:** STM and LEED analysis of the surface superstructure. a,b) Atomically resolved STM constant-height images of the H-phase obtained with two different contrasts at 0.2 V and 0.6 V, average currents 2.6 nA and 4.5 nA, respectively. The white rhombus tentatively marks the superstructure unit cell. c) Detail of the LEED pattern, composed of two images (red and green), taken on adjacent domains at 30 eV with 185 nm electron beam spot diameter. The orange and cyan vectors mark the periodicities corresponding to the bulk α-$Fe_2O_3$(0001) and top layer, respectively. The black vector and the dashed line mark the periodicity and orientation of one moiré domain.

The overall character of the LEED pattern and the contrast in the STM images is consistent with a moiré arising from the coincidence of two unequal lattices; given the well-defined bulk properties, this can only be explained by the existence of an Fe oxide layer with specific lattice constant and atomic structure, different from its carrier - the α-$Fe_2O_3$(0001) bulk crystal.

However, the composition of the top layer and exact atomic arrangement remains ambiguous. To address this ambiguity we have performed low-energy ion scattering (LEIS) experiments employing $He^+$ ions with energies of 1.4 - 5.0 keV. LEIS is primarily sensitive to composition of the topmost atomic layer making it particularly useful to determine the surface termination. The scattering spectra measured on the H-phase given in Fig. S2 show the presence of both Fe and O species on the surface (a comprehensive description of the measurements and quantification is given in the SI). From the peak intensities and energy dependence we have determined the Fe and O atom concentrations as $0.5 \times 10^{15}$ $cm^{-2}$ and $1.7 \times 10^{15}$ $cm^{-2}$, respectively, which indicates prevalence of oxygen in the topmost layer and in the near surface region accessible by LEIS.

The abundance of oxygen in the terminal layers narrows the choice of possible models explaining the surface termination. We design previously unexplored models that can potentially be self-supporting and not disrupted by the substrate. Promising candidates are layered structures analogous to transition metal dichalcogenides, in which a hexagonal lattice of Fe atoms is sandwiched between two hexagonal O layers. Two most common structures of transition metal dichalcogenides are a hexagonal 2H phase (space group, P63/mmc) and a trigonal 1T phase (space group, P$\bar{3}$m1), shown in Fig. S4. To the best of our knowledge, such structures of an iron oxide have been neither proposed nor observed so far. Therefore, we employ first-principles DFT calculations to elucidate the basic properties of this form of iron oxide. The calculated phonon spectra of the single-layer free-standing 2H-$FeO_2$ and 1T-$FeO_2$ (Fig. S5) show that both structures are dynamically stable, i.e. there are no imaginary frequencies which would indicate a spontaneous collapse. The total energy of the 1T phase is lower by 0.88 eV per formula unit than that of the 2H, thus the 1T phase is thermodynamically far more stable (for details about how the calculations were carried out and the stability of the models please see SI).

Having established through simulation that 1T-FeO$_2$ is stable and thermodynamically favourable, we investigate its magnetic and electronic properties by evaluating the energy of several possible magnetic states of this structure: high-spin ferromagnetic configuration, low-spin ferromagnetic (FM), and antiferromagnetic (AFM) alignment of spins. The spin moments in each configuration reside on Fe atoms, while the projected spin moments on O atoms are negligible. The most stable configuration turns out to be AFM; the energy of the high spin FM ordering is 0.14 eV per formula unit higher. The low spin FM is an additional 0.97 eV higher than the FM value. The projected spin moments are ±3.79 $\mu_B$ in the AFM configuration. The spin moments are parallel in one of the in-plane directions and antiparallel in the second. The lattice parameter of the free standing 1T-FeO$_2$ is 2.97 Å in the AFM state and 2.98 Å in the high-spin FM state. The projection of the density of states reveals that the free-standing 1T-FeO$_2$ is metallic (Fig. S6) in all spin configurations. For the supported 1T-FeO$_2$, the coupling to underlying hematite support can influence the position of its Fermi level and hence the layer conductivity and spin configuration.

Finally, we shall address how the layer is bonded to the hematite bulk. Its structure consists of separate hexagonal Fe and O$_3$ layers stacked as a sequence of Fe-O$_3$-Fe units along the [0001] direction, meaning that the unreconstructed (0001) surface of hematite has three possible terminations. Earlier DFT calculations suggested that the Fe termination containing half of the inter-plane Fe, termed a "half-metal" termination, should be stable at low oxygen pressures [42]. According to the same work the O$_3$-termination becomes stabilized at high oxygen pressure by a substantial relaxation of the surface atoms.

Two full supercell models were used, made of $\begin{smallmatrix}13 & 1\\ -1 & 12\end{smallmatrix}$ 1T-FeO$_2$ relaxed on $\begin{smallmatrix}9 & 5\\ -5 & 4\end{smallmatrix}$ Fe- and O$_3$-terminated α-Fe$_2$O$_3$(0001), which matched best with the geometrical parameters of the real system (details about the geometrical derivation of the model from experimental data can be found in the SI and Fig. S7. Corresponding profiles of the atomic models are shown in Fig. 4b and S8, respectively). The resulting calculated formation energies, displayed in Fig. S9, demonstrate that the model of the 1T-FeO$_2$ on the Fe-terminated hematite is more stable than O$_3$-terminated hematite under any reasonable chemical potential of oxygen. The stability is due to stronger bonding of 1T-FeO$_2$ to the support. The estimation of the binding energy ( energy needed to cleave the FeO$_2$ layer from the support) is 1.07 J/m$^2$ for the model on Fe-terminated bulk, whereas it amounts to just 0.12 J/m$^2$ for the O$_3$-terminated support, a value typical for noncovalent van der Waals bonding.

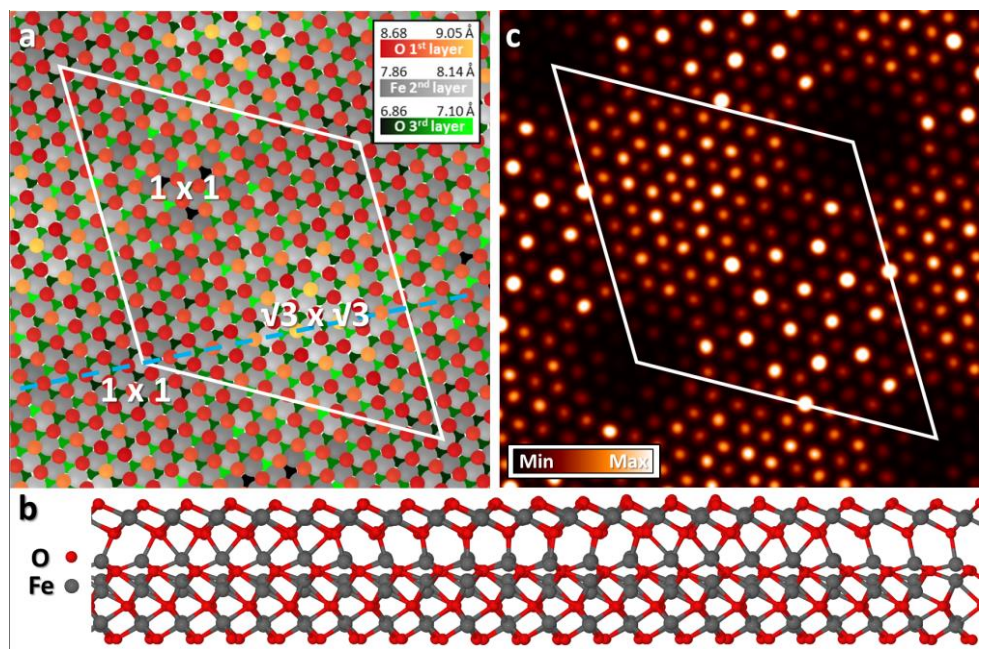

**Fig. 4:** Evaluation of the 1T-FeO$_2$ on α-Fe$_2$O$_3$(0001) relaxed model. a) the scheme of atomic heights in the three topmost surface layers (O 1$^{st}$, Fe 2$^{nd}$ and O 3$^{rd}$ layers, with the unit cell marked by a white rhombus, and local reconstructions labelled, b) ball-and-stick profile of the structure. The plane of projection is denoted by the blue dashed line in a). c) simulation of a constant-height STM image.

The interface Fe atom coordination geometry varies gradually throughout the supercell between tetra- and octahedral, due to the lattice mismatch with the O atoms in the 3$^{rd}$ layer, and leads to a relaxation of the O and Fe atoms in the entire FeO$_2$ layer in directions both parallel and vertical with respect to the surface. Moiré modulation and local √3×√3 periodicity arise within the FeO$_2$ layer, as a consequence of the height variation of the topmost O atoms (1$^{st}$ layer), as depicted by the model in Fig. 4a, color coded according to separate layers and individual atom heights. This phenomenon effectively splits the supercell into three dissimilar regions, of which one has √3×√3 and two retain the original 1×1 periodicity. The height variation of the O atoms in the 1st layer directly translates into the contrast variation obtained by a constant-height STM simulation, in Fig. 4c, made with the Tersoff-Hamann approximation, at a distance of 0.25 nm from the surface plane, tunneling in the range -0.5 eV to 0 eV. The resulting image yields the contrast and characteristic features consistent with both the original [19,20] and present STM observations. The projection of spin moments indicates a local rearrangement of the Fe electrons within the 1×1 regions; the projected spin values for Fe atoms vary locally between 2.6 and 4.2 µB in the dark 1x1 area and between 3.3 to 4.4 µB in the brighter 1×1 area. In the √3×√3 regions, the spin is distributed more evenly - between 3.6 and 4.3 µB.

**Conclusions**

We conclude from our experimental evidence, well supported by fully relaxed large scale calculations, that the surface termination of α-Fe$_2$O$_3$(0001) takes the form of a continuous two-dimensional 1T-FeO$_2$ layer attached to the half-metal termination of the bulk crystal. The proposed model matches the available experimental evidence and resolves the long-standing controversy about the termination of the α-Fe$_2$O$_3$(0001) surface. According to the theoretical calculations, the 2D layer of 1T-FeO$_2$ possesses interesting material properties, a metallic character with antiferromagnetic spin arrangement, highly appealing for spintronic applications. We anticipate that these findings can stimulate new research in the field of 2D oxide materials.

**Acknowledgements**

We would like to thank to Prof. H. J. Freund (Fritz-Haber Institute, Berlin), Dr. Jaeyoung Kim (PAL, Pohang, Korea) for fruitful discussions and support and Lukáš Kormoš for assistance with sample preparation. The research was supported by the Ministry of Education, Youth and Sports of the Czech Republic LQ1601 (CEITEC 2020). Part of the work was carried out with the support of CEITEC Nano Research Infrastructure (MEYS CR, 2016–2019). The authors also acknowledge the CERIC-ERIC Consortium for the access to experimental facilities. The authors are grateful to Jack Hellerstedt for assistance in writing of the manuscript.

**Author Contributions**

J.R performed the main STM measurements. P.P. and J.Č. realized the LEEM experiments. A.C. and B.M. developed the preparation procedure. S.P. carried out the LEIS measurements and data analysis. B.Š. carried out the XPS measurements. J.B. assisted in preparation of the samples. P.L. performed the theoretical calculations, J.L. assisted in the creation of the geometrical model. M.Š. and J.R. evaluated the experimental data and wrote the paper. M.Š. and P.J. proposed and initiated the experiments and calculations. All the authors reviewed and discussed the manuscript thoroughly before submission.

**Methods**

Natural α-Fe$_2$O$_3$(0001) single crystal samples were acquired from SurfaceNet GmbH and mounted on Mo/Ta plates for STM, LEED and XPS measurements. STM and LEED characterization was carried out at room temperature in a UHV VT-STM (Scienta Omicron GmbH), XPS spectra were recorded at the NAP-XPS (Specs GmbH) of the Faculty of Mathematics and Physics of the Charles University. The temperature was monitored by a Micro-Epsilon ® 2MH - CF3 thermoMETER of 1.6 µm spectral range focused on the center of the sample, measuring in the range of 385-1600 ºC with and spectral emissivity of 0.6. Electrochemically etched W tips were used for scanning. LEEM/µLEED and LEIS measurements were performed with a FE-LEEM P90 (Specs GmbH) instrument and a Qtac 100 (Ion TOF GmbH), respectively. Here the samples were prepared in a separate UHV chamber and moved to the instruments for analysis via a linear transfer system without breaking the UHV conditions; the base pressure in all chambers was below 2×10$^{-10}$ mbar. In the preparation done for LEEM and LEIS, the sample temperature was calibrated by a K-type thermocouple and checked by the pyrometer.

**Data Availability**

The data that support the findings of this study are available from the corresponding authors upon reasonable request.

**Identification of two-dimensional FeO$_2$ termination of hematite α-Fe$_2$O$_3$(0001) surface - Supporting Information**

*Jesús Redondo*[1,2], Petr Lazar[3], Pavel Procházka[4], Stanislav Průša[4], Jan Lachnitt[2], Benjamín Mallada[1], Aleš Cahlík[1], Jan Berger[1], Břetislav Šmíd[4], Pavel Jelínek[1], Jan Čechal*[4], Martin Švec*[1]*

1 - Institute of Physics, Czech Academy of Sciences, Prague, Czech Republic
2 - Faculty of Mathematics and Physics, Charles University, Prague, Czech Republic
3 - Regional Center for Advanced Materials and Technologies, Olomouc, Czech Republic
4 - Central European Institute of Technology, Brno, Czech Republic

**Effect of air exposure and annealing in vacuum**

All the α-Fe$_2$O$_3$(0001) samples were mechanically cut and polished from natural hematite crystals. The characteristic LEED pattern of a mixture of R- and H- phases is obtained upon degassing as received samples in UHV at 700 °C, and the surface is clean enough to achieve atomic resolution in STM.

In order to study the effects of air exposure on the H- phase, we prepared sample completely covered with the H-phase. A blurry LEED pattern is observed after 30 min air exposure. Annealing of the air-exposed H- phase leads to a partial transformation of the H-phase to the R-phase, as evidenced in Fig. S1a. STM image indicates that ~ 30% of the H-phase is reduced.

XPS measurements carried out after air exposure before annealing reveal the presence of adventitious carbon on the surface, as shown in blue in Fig S1b. The four main sources of carbon contamination of surfaces exposed to air and UHV are graphite, hydrocarbons, CO and CO$_2$. In the case of iron and iron oxides surfaces, it has been proposed that the source of carbon is mainly CO and CO$_2$, which are fixed to the surface and form a thicker layer on the surface the longer the exposition time [1]. For this reason, we relate the partial decomposition of the H-phase to the reaction with the adsorbed C species.

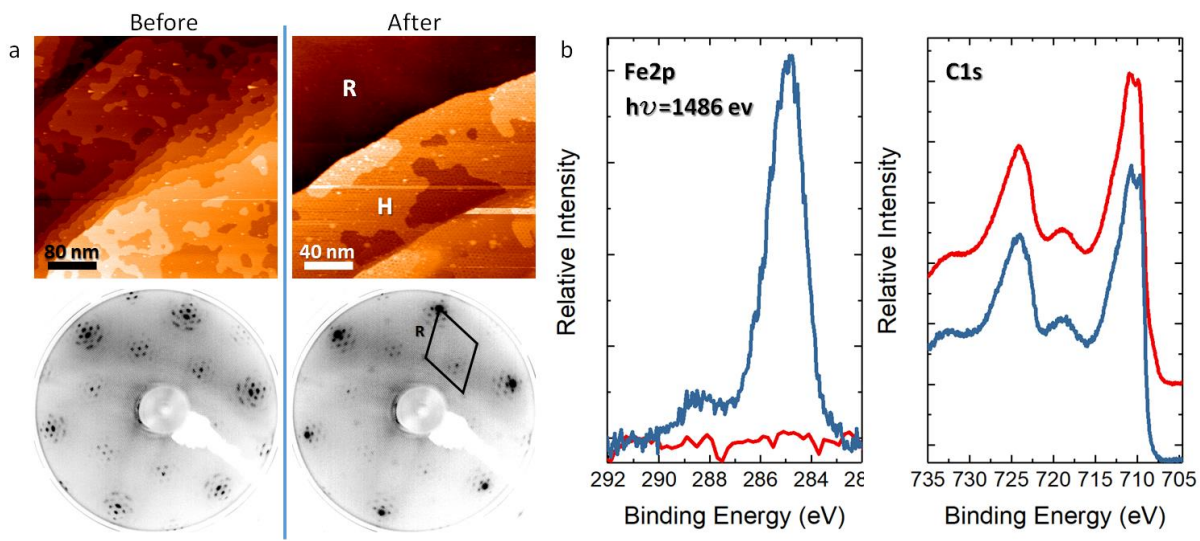

**Fig.S1**: Comparison of the α-Fe$_2$O$_3$(0001) before and after exposure to ambient air conditions. a) STM and LEED characterization of the surface. A portion of the H-phase transforms upon annealing. b) XPS spectra of the C1s and Fe 2p peaks before (red) and after (blue) air exposure (without annealing).

Finally, we investigated the results of annealing both pure R- and H-phases up to 900 ºC for 18 hours in UHV (1×10$^{-9}$ mbar). Based on LEED and STM measurements, we conclude that the surface structures and total coverages of the R- and H-phase remain unaltered, contrary to the transformations from H- to R- phase and vice versa observed on thin α-Fe$_2$O$_3$(0001) films grown metals [2].

**Low-energy ion scattering experiments**

The LEIS technique is known for its extreme surface sensitivity[3]. The elemental composition of the top-most atomic layer is reflected by the presence of surface peaks of individual elements in the measured energy spectrum. The signal intensity (surface peak area) is proportional to the surface atomic concentration of the respective element $n_i$ [4]:

$$S_i = n_i \frac{d\sigma_i}{d\Omega} c P_i^+.$$

Here the measured signal is normalized to the unit primary beam charge of 1 nC (see the vertical scale unit in Fig. S2).

The differential scattering cross section $d\sigma/d\Omega$ is followed by the instrumental factor $c$, and $P^+$ is the ion fraction representing the probability that the projectile leaves the surface in the charged state. The ion fraction $P^+$ is expressed in the form of an exponential function, in which $v_c$ is the characteristic velocity of the projectile–target atom pair [3]:

$$P^+ = exp\left(v_c \frac{1}{v}\right).$$

$\frac{1}{v}$ expresses the sum of $\frac{1}{v_i}$ and $\frac{1}{v_f}$, where $v_i$ and $v_f$ are the limit velocities of the projectile at incoming and outgoing trajectories. The particular definition of the inverse velocities depends on the actual combination of the charge exchange processes involved in the collision.

The theoretical prediction of $P^+$ is difficult. The ion fraction is influenced by many charge exchange processes between the projectile and the surface under analysis. Each neutralization mechanism is dominant at a specific distance between the projectile and the target surface or an individual target atom. We will focus on two charge exchange processes that are relevant to our analysis; the Auger neutralization (AN) and the collision induced neutralization (CIN). While AN is effective when the projectile and target atom orbitals overlap, CIN dominates during the closest approach of the colliding partners. Thus, AN defines the inverse velocity at lower primary energies, and CIN take over at the higher primary energies.

Figure S2 contains four spectra measured under identical experimental conditions (helium primary beam with kinetic energy 3.0 keV and scattering angle 145°). The red spectrum represents the signal measured on the H-phase of α-$Fe_2O_3$(0001); it features an Fe surface peak at 2310 eV and an O peak at 1180 eV. The green and black spectra were measured on polycrystalline Cu and Fe references cleaned by Ar sputtering (fluence 3×10$^{16}$ ions/cm²). The blue spectrum was measured on a $SiO_2$ surface cleaned by Ne sputtering (fluence 1×10$^{16}$ ions/cm²).

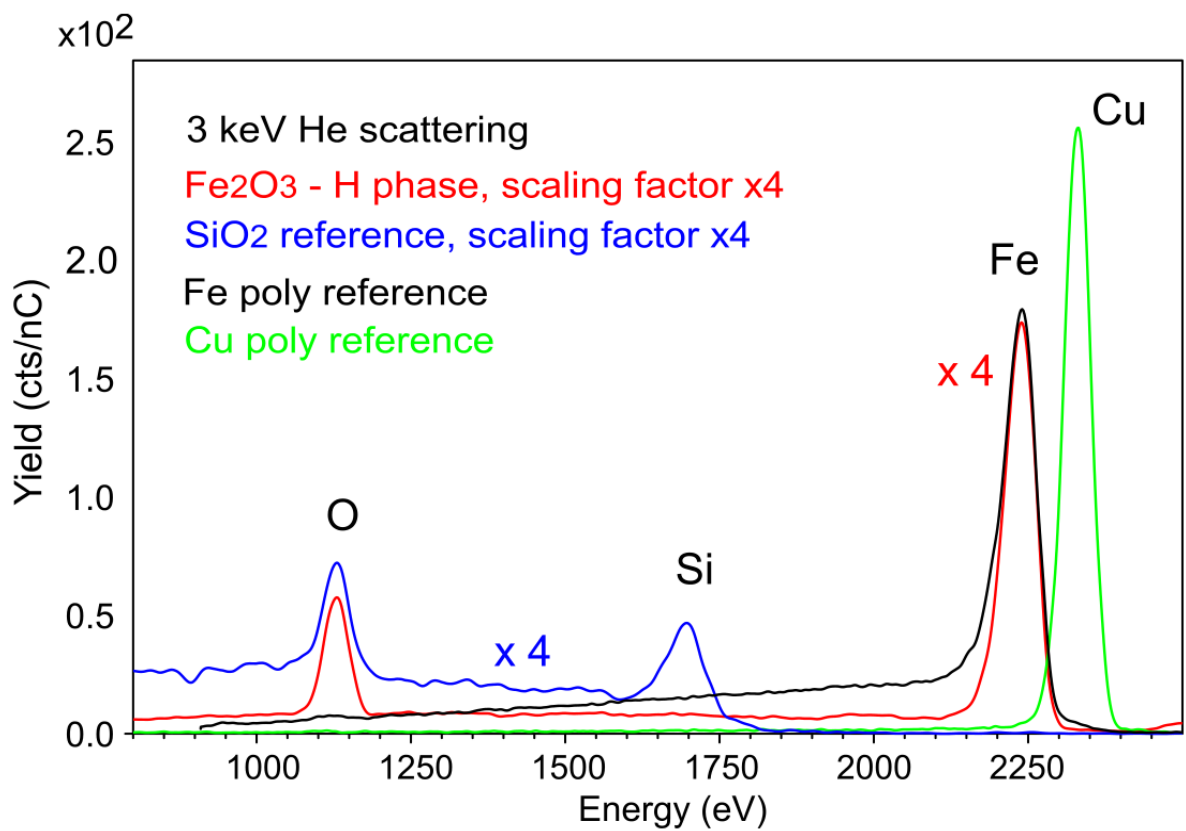

**Fig.S2:** LEIS spectra for helium (3.0 keV) scattering (145°) on the analyzed α-Fe$_2$O$_3$(0001) surface (H-phase) and on references (Fe, SiO$_2$, and Cu). The spectra for SiO$_2$ and Fe$_2$O$_3$ were multiplied by a scaling factor of 4 for better visibility.

The instrumental factor $c$ was evaluated from the helium scattering on a clean Cu polycrystalline surface within the range of primary energies from 1.0 keV to 2.2 keV. It is the AN mechanism which is dominant in this case, and thus the perpendicular inverse velocities were used in the calculation [3].

The iron atomic surface concentration is evaluated using signals of relevant Fe peaks in Fig. S2 (primary energy 3.0 keV). Apparently, the charge exchange processes for the measured surface and Fe polycrystalline reference are dominated by CIN. The ratio is equal to 0.24 (please mind the scaling factor of 4 in Fig. S2). After the multiplication by the atomic surface concentration for the Fe reference 1.93×10$^{15}$ cm$^{-2}$ (calculated from the Fe density 7.87 g/cm$^3$) we obtain the value 0.46×10$^{15}$ cm$^{-2}$ for the H-phase.

Quantification of the O content is more demanding, because both neutralization mechanisms are involved within the energy range used for our LEIS analysis; AN being dominant at lower energies and CIN at higher energies. These two energy ranges were distinguished for helium scattering on the $SiO_2$ surface by Téllez et al. [5]. The CIN becomes dominant for primary energies above 3 keV.

Signal intensities divided by the instrumental factor and by the differential scattering cross-section are plotted in logarithmic scale against the sum of inverse velocities in Fig. S3. The ratio of intercepts of the fitted straight lines with the vertical axis (at infinite velocity) determines the ratio of O atomic surface densities in the analyzed surface and in the silica reference:

$$\frac{N_O^{Fe2O3}}{N_O^{SiO2}} = \frac{35.7}{34.7}$$

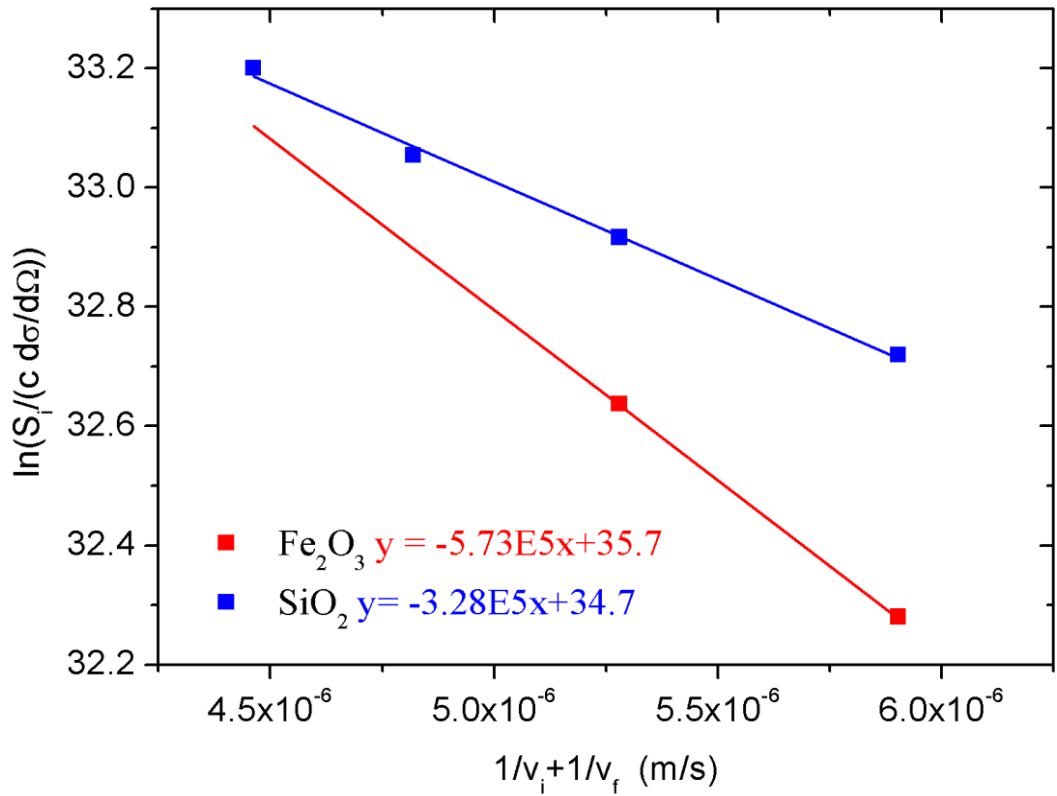

**Fig.S3:** Logarithmic plot of the O signal divided by the instrumental factor and by the scattering cross section for the analyzed surface (red) and for the silica reference (blue).

Taking into account the atomic surface concentration of O in the reference ($N_O^{SiO2}$ = 1.63×10$^{15}$ cm$^{-2}$ calculated from the $SiO_2$ density 2.32 g/cm³) we obtain the sought concentration $N_O^{Fe2O3}$ = 1.68×10$^{15}$ cm$^{-2}$ for the H-phase.

The calculated atomic surface concentration of O is 3.6 times higher than that of Fe atoms. It has to be mentioned that the O evaluation on H-phase is based on only two experimental points (primary energies 4.0 keV and 5.0 keV) and thus the estimated relative uncertainty in the atomic surface concentration is about 30%. In summary, the LEIS analysis of the H-phase of the α-$Fe_2O_3$(0001) surface shows that oxygen atoms clearly predominate in the topmost atomic layer.

Nevertheless, the large difference of the atomic surface concentration of Fe and O in the $Fe_2O_3$ H-phase should be explained. The proposed 1T-$FeO_2$ model is terminated by an O atomic layer situated (after relaxation) about 0.9 Å above the plane with Fe atoms. This arrangement would exhibit dominant scattering of the O atoms over that of the Fe atoms.

The calculation based on screened Coulomb potential [6] shows that the shadow cone radius for a 3.0-keV He projectile has at the distance of 0.9 Å behind the O atom a radius of only 0.27 Å. Thus, the shadow cones formed by O atoms at the incoming trajectory are not able to effectively screen the Fe atoms.

The situation is different with the outgoing trajectory, where the helium projectiles in specific azimuths pass relatively close to the upper O atoms (the Qtac 100 instrument collects scattered ions over the whole azimuth to increase the detection sensitivity). The resulting Fe signal is reduced by shielding due to neutralization at the vicinity of the O atoms. This mechanism was described and experimentally proved by van den Oetelaar et al.[6] for He scattering on polycrystalline CuO. The Cu signal was reduced 5 times with respect to a signal estimated for Cu concentration calculated from the bulk density of CuO.

**Theoretical calculations**

The projector-augmented wave method, as implemented in the Vienna ab-initio simulation package (VASP) suite [7,8], was used for the calculations. The energy cutoff for the plane-wave expansion was set to 400 eV. For the exchange-correlation functional, we chose a semi-local generalized gradient approximation within the PBE parameterization [9] in combination with effective Hubbard hamiltonian which described on-site Coulomb repulsion among d-electrons by an additional term U. Transition metal oxides are well known as strongly correlated materials due to repulsive electron-electron interaction between d-electrons and classical DFT functionals typically fail to describe such systems. We adopted simplified rotationally invariant approach to the GGA+U introduced by Dudarev *et al.* [9,10] and set $U_{eff}$ = U-J parameter to 5.3 eV[11,12]. We used also hybrid screened exchange functional HSE06 [13] to test the results for free-standing $FeO_2$. For the simulations of STM, we used a standard Tersoff-Hamann method, integrating the tunneling in the energy range -0.5 to 0.0 eV.

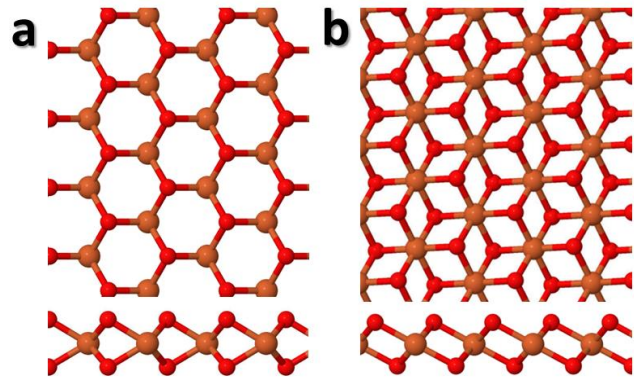

**Fig.S4:** The models of free-standing a) Hexagonal 2H form and b) trigonal 1T form of the $FeO_2$ layer.

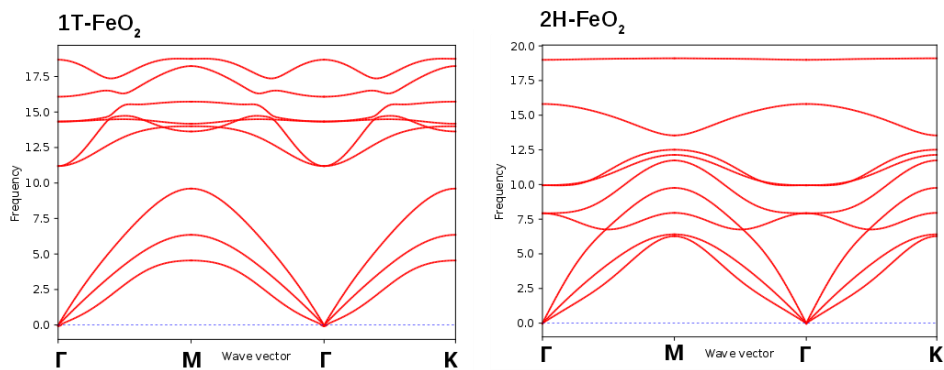

**Fig.S5:** Phonon dispersions (phonon frequency in THz) of free-standing single-layer of 1T-$FeO_2$ and 2H-$FeO_2$ calculated via density functional perturbation theory approach.

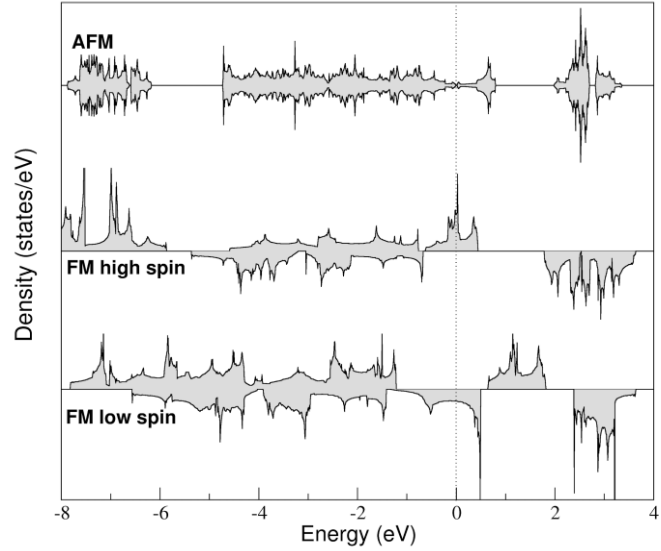

**Fig.S6:** The electronic density of states for the antiferromagnetic (AFM), high spin ferromagnetic (FM), and low spin FM phase of 1T-FeO$_2$. The Fermi level is aligned to zero of the energy.

**Design of the full supercell models**

For the calculation of the top lattice parameter ($a_{top}$), we use the known bulk hematite lattice constant $a_{bulk}$ = 5.038 ± 0.002 Å [14], the superstructure lattice parameter $a_{moiré}$ = 40 ± 1 Å measured by STM, and the relative superstructure orientation between domains $2\varphi$ = 8° ± 1°. Assuming that the superstructure arises from the interference of two lattices (a moiré), we have calculated the top layer lattice parameter $a_{top}$ = 3.14 ± 0.01 Å, and the lattice rotation with respect to the bulk lattice 30° ± δ (for the two domains) with δ = 0.31° ± 0.04°. The top layer lattice parameter is within the range determined by the STM measurements (3.08 ± 0.08 Å).

The obtained values describe a possibly incommensurate surface structure, and STM confirms that the superstructure is indeed incommensurate. For the purpose of the calculations, we have designed a commensurate model that best matches the measured parameters. The model is shown in Fig. S7, and it comprises $\begin{pmatrix} 13 & 1 \\ -1 & 12 \end{pmatrix}$ top-layer cells on $\begin{pmatrix} 9 & 5 \\ -5 & 4 \end{pmatrix}$ bulk cells or, in Wood's notation, a single-layer 1T-FeO$_2$-($\sqrt{157} \times \sqrt{157}$)R−4.0 on α-Fe$_2$O$_3$(0001)-($\sqrt{61} \times \sqrt{61}$)-R26.3 (for the 30° + δ domain). The parameters of this model are $a_{moiré,model}$ = 39.35 Å, $a_{top,model}$ = 3.14 Å, $δ_{model}$ = 0.29°, and $2\varphi_{model}$ = 7.34°, in excellent agreement with the experimental values.

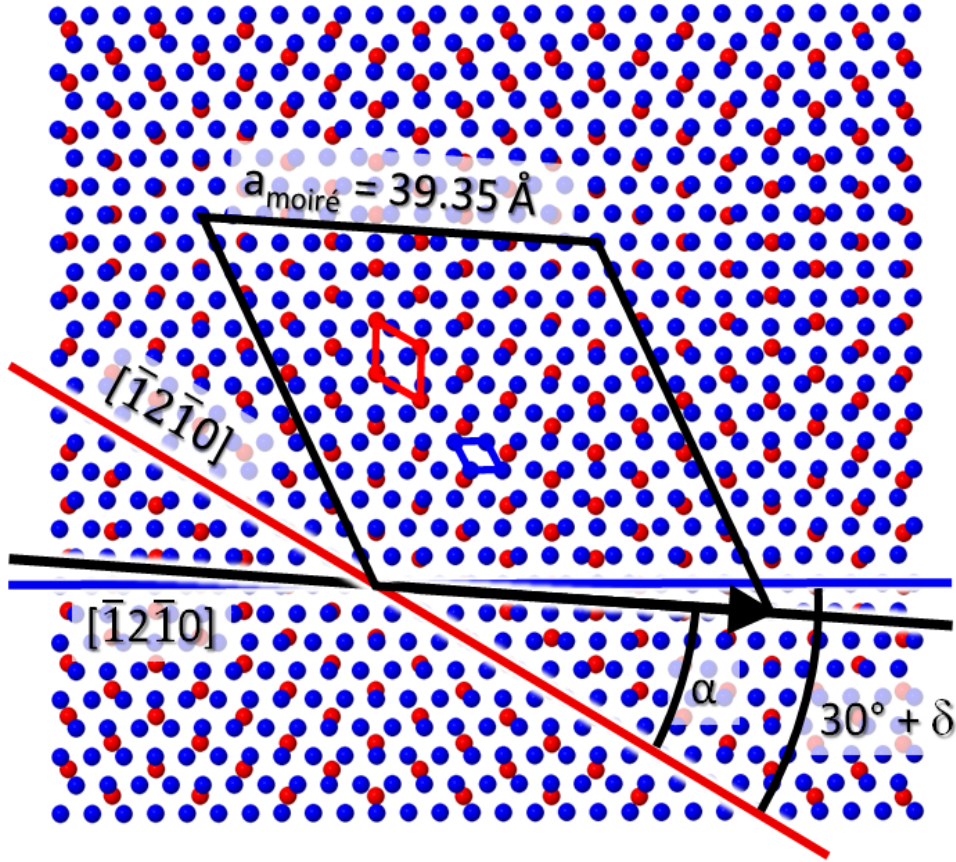

**Fig. S7:** One of the mirror-symmetrical variants of the best-matching Moire model. Hexagonal grid representing the overlayer periodicity (3.14 Å blue dots) is placed on the top of a grid with the α-Fe$_2$O$_3$(0001) parameters (5.039 Å, red dots).

**Stability of the models**

Thermodynamic stability of various forms of an Fe oxide overlayer and hematite surface terminations can be assessed from the formation energy $E_{form}$

$$E_{form} = E_{tot} - n_{Fe}\, \mu_{Fe} - n_O\, \mu_O$$

where $E_{tot}$ is the total energy of the system, $\mu_{Fe}$ the chemical potential of Fe, $\mu_O$ the chemical potential of O, and $n_{Fe}$, $n_O$ the respective number of atoms. The temperature and pressure dependence of the formation energy was omitted as it makes negligible contribution at our experimental conditions. The Fe and O chemical potential are not independent, they are bound together by the existence of the hematite bulk phase. That allows to express $E_{form}$ as a function of the O chemical potential

$$E_{form} = E_{tot} - \tfrac{1}{2}\, n_{Fe}\, \mu_{Fe2O3} - (3/2\ n_{Fe} - n_O)\, \mu_O$$

The meaningful range of $\mu_O$ is limited by the conditions that the chemical potential of Fe has to be smaller than that of an atom of bulk Fe, and that the chemical potential of O has to be smaller than that of an O atom of $O_2$ molecule.

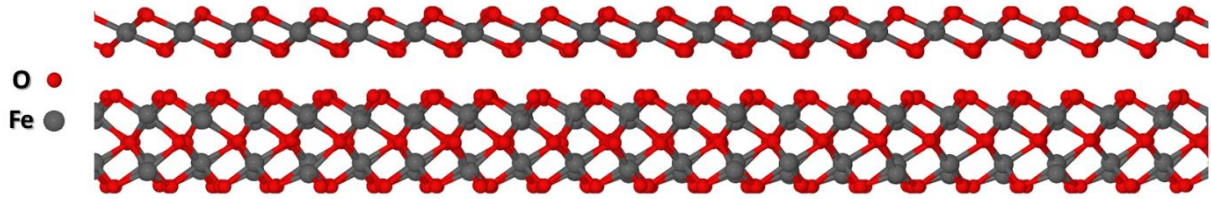

**Fig. S8:** Side-view of the ball-and-stick model of 1T-$FeO_2$ on O3-terminated α-$Fe_2O_3$(0001) bulk

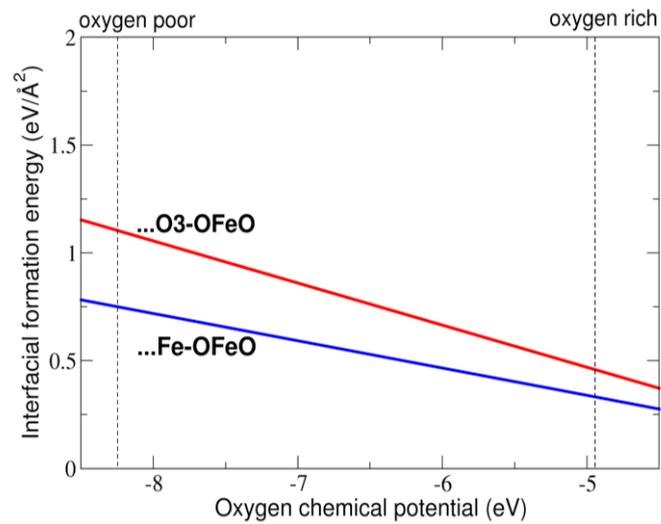

**Fig. S9:** Formation energies of the $FeO_2$-hematite interface for two terminations of the bulk hematite, the half-metal (Fe) termination and the $O_3$ termination, as functions of the oxygen chemical potential, $\mu_O$ (O partial pressure). The physically possible range of $\mu_O$ is marked by vertical dashed lines).